\documentclass[12pt,preprint]{aastex}
\usepackage{epsfig}

\def\arcmin{^{\prime}}

\def\gtrsim{\mathrel{\hbox{\rlap{\hbox{\lower4pt\hbox{$\sim$}}}\hbox{$>$}}}}
\def\lesssim{\mathrel{\hbox{\rlap{\hbox{\lower4pt\hbox{$\sim$}}}\hbox{$<$}}}}

\def\vkm{km s$^{-1}$}

\def\degree{$^\circ$}

\def\arcsa#1#2{$#1^{\prime\prime}_{^\textrm{.}}#2$}

\def\solarmass{$M_\odot$}
\def\solarmasspy{$M_\odot$ yr$^{-1}$}

\def\mJyb{mJy beam$^{-1}$}
\def\uJyb{$\mu$Jy beam$^{-1}$}

\def\cmc{cm$^{-3}$}
\def\cms{cm$^{-2}$}
\def\degr{\hbox{$^\circ$}}
\def\micron{$\mu$m}
\def\VLSR{V_\textrm{\scriptsize LSR}}
\def\Vsys{V_\textrm{\scriptsize sys}}
\def\Voff{V_\textrm{\scriptsize off}}

\def\ra#1#2#3#4{#1^\mathrm{h} #2^\mathrm{m} #3^\mathrm{s}_{^\textrm{.}} #4}
\def\dec#1#2#3#4{#1\degr #2\arcmin #3^{\prime\prime}_{^\textrm{.}}#4}

\def\mH2{m_{\textrm{\scriptsize H}_2}}
\def\Ro{R_\textrm{\scriptsize 0}}

\def\To{T_\textrm{\scriptsize 0}}
\def\no{n_\textrm{\scriptsize 0}}

\def\H2{H$_2$}
\def\N2HP{N$_2$H$^+$}

\def\NH3{NH$_3$}

\def\putfig#1#2#3{\epsfig{scale=#1,angle=#2,figure=#3}}
\def\leftblank#1{}

\newcounter{mfigure}[section]

\begin{document}


\title{First Detection of A Linear Structure in the Midplane of the Young HH
211 Protostellar Disk: A Spiral Arm?}


\author{Chin-Fei Lee\altaffilmark{1,2}, Kai-Syun Jhan\altaffilmark{1,2}, and Anthony Moraghan\altaffilmark{1} }
\altaffiltext{1}{Academia Sinica Institute of Astronomy and Astrophysics,
P.O. Box 23-141, Taipei 106, Taiwan; cflee@asiaa.sinica.edu.tw}
\altaffiltext{2}{Graduate Institute of Astronomy and Astrophysics, National Taiwan
   University, No.  1, Sec.  4, Roosevelt Road, Taipei 10617, Taiwan}

\begin{abstract}






Spiral structures have been detected in evolved protostellar disks, driving
the disk accretion towards the central protostars to facilitate star
formation.  However, it is still unclear if these structures can form
earlier in young protostellar disks.  With the Atacama Large
Millimeter/submillimeter Array (ALMA), we have detected and spatially
resolved a very young and nearly edge-on dusty disk with a radius of only
$\sim$ 20 au in the HH 211 protostellar system at submillimeter wavelength. 
It is geometrically thick, indicating that the submillimeter light emitting
dust grains have yet to settle to the midplane for planet formation. 
Intriguingly, it shows 3 bright linear structures parallel to the equatorial
plane, resembling a 3-layer pancake that has not been seen before.  The top
and bottom ones arise from the warm disk surfaces, unveiling the flared
structure of the disk.  More importantly, the middle one is in the dense
midplane of the disk and can be modeled as a trailing spiral arm excited by
disk gravity, as seen in evolved protostellar disks, supporting the presence
of spiral structures in the very early phase for disk accretion.




\end{abstract}


\section{Introduction}

Protostellar disks play a crucial role in forming extrasolar systems, as
they not only facilitate the growth of central protostars into mature stars
\citep{Shu1987}, but also evolve into protoplanetary disks which serve as
the primary site for planet formation
\citep{Williams2011,Brogan2015,Andrews2020}.  They form because the
gravitationally collapsing material carries part, if not all, of the angular
momentum toward the center.  In the early phase of star formation, the disks
are small and deeply embedded in dusty molecular cores, and thus are hard to
be detected and spatially resolved.  In addition, magnetic braking in some
collapsing models can remove the angular momentum from the collapsing
material efficiently and thus prevent the disks from forming \citep{Allen2003}. 
Thus, it is still unclear how early they can form and what their initial
structures are.  It is also unclear what mechanisms can transport angular
momentum within the disks or remove it from the disks so that the disk
material can flow in from the outer to the inner part and then feed the
central protostars.  Possible mechanisms include spiral arms induced by
gravitational instability (GI) \citep{Bate1998,Kratter2016}, disk-driven
winds \citep{Konigl2000,Turner2014}, and magneto-rotational instability
\citep{Balbus2003,Turner2014}.

The HH 211 protostellar system is one of the youngest systems
\citep{Froebrich2005}, allowing us to address above questions.  It is
located in Perseus at $\sim$ 321 pc away.  Its formation started with a
gravitational collapse only $\sim$ 35000 yrs ago and the central protostar
has a mass of $<$ 0.08 \solarmass{} \citep{Lee2018HH211,Lee2019HH211}.  It
is associated with a prominent jet
\citep{McCaughrean1994,Gueth1999,Hirano2006,Jhan2021}, indicating that a
disk must have formed and is actively accreting onto the protostar. 
Previous studies have suggested the presence of a dusty disk around the
protostar, which is likely nearly edge-on and spatially unresolved
\citep{Segura-Cox2016,Lee2018HH211}. Here we
report the detailed structure of the disk resolved with our ALMA observation
in dust continuum at $\sim$ 4 au resolution, not only unveiling the earliest
structure of a disk but also challenging the theories of disk formation
during such an early phase.  We also derive the physical properties of the
disk and discuss how the observed structure in the disk can be formed and
what can drive the accretion process.


\section{Observations} \label{sec:obs}

The HH 211 protostellar system was mapped with ALMA in Band 7 with one
pointing toward the center in Cycle 7 (Project ID: 2019.1.00570.S).  It was
mapped with 4 executions in 2021, with one on August 29, two on September
27, and one on September 29.  It was observed for a total time of $\sim$ 167
minutes using 44-45 antennas and configurations of C43-9 and C43-10 with
baseline lengths of 70 m --16.2 km.  The maximum recoverable scale was
$\sim$ \arcsa{0}{3}.  The correlator was setup to have 6 spectral windows,
with 4 having a bandwidth of 0.234 GHz and a velocity resolution of 0.2
\vkm{} per channel, one having a bandwidth of 0.469 GHz and a velocity
resolution of 0.4 \vkm{} per channel, and one having a bandwidth of 1.875
GHz and a velocity resolution of 0.8 \vkm{} per channel.  Since we focus on
the dusty disk, line-free channels were used to derive the continuum with a
bandwidth of $\sim$ 3.3 GHz centered at $\sim$ 352 GHz.


The CASA package 6.2.1.7 was used to calibrate the data.  Quasars J0238+1636
and J0510+1800 were used as both the bandpass and flux calibrators.  Quasar
J0336+3218 was used as a phase calibrator.  Line-free channels were combined
to produce the continuum channels.  Robust weighting factors of 2 and 0 were
used for the visibility to generate the continuum maps at $\sim$
\arcsa{0}{040}$\times$\arcsa{0}{025} resolution with a noise level of $\sim$
55 \uJyb{} (or 0.54 K) and \arcsa{0}{022}$\times$\arcsa{0}{013} resolution
with a noise level of $\sim$ 46 \uJyb{} (or 1.6 K), respectively, in order
to study the envelope and disk.  The 4th spectral window also contained the
SiO J=8-7 line that was used to map the jet in order to determine the jet
axis.  A robust factor of 0.5 was used to make the SiO maps with an angular
resolution of \arcsa{0}{029}$\times$\arcsa{0}{018}.  The SiO maps have
a channel width of 0.42 \vkm{} and a RMS of $\sim$ 2 \mJyb{} for each
channel.

\section{Results}

\subsection{Inner Envelope and Disk}

Figure \ref{fig:cont}a presents the continuum emission intensity map within
$\sim$ 100 au of the central protostar, observed at the frequency of $\sim$
352 GHz at a resolution of 12.8 au$\times$8.0 au (or
\arcsa{0}{040}$\times$\arcsa{0}{025}). At this frequency, the continuum
emission is the thermal emission arising from the dust, based on the SED
analysis \citep{Lee2007HH211}.  The region outlined by the contours extends
to the northeast (NE) and southwest (SW) from a compact central region,
tracing the inner part of the flattened envelope currently collapsing
(infalling) with rotation toward the center \citep{Lee2019HH211}.  The
contours are crowded around the compact central region, indicative of a
rapid increase in the emission intensity there, supporting a structural
change from a tenuous envelope to a dense disk.
Zooming into the compact central region at $\sim$ 2 times higher resolution
of 7.1 au$\times$4.1 au (or \arcsa{0}{022}$\times$\arcsa{0}{013}) as shown
in Figure \ref{fig:cont}b, we detect a disk and spatially resolve it in both
the equatorial and vertical directions.  The disk is oriented perpendicular
to the SiO jet axis (as indicated by the blue and red arrows, see Figure
\ref{fig:cont}b). 

\subsection{SiO Jet Axis}

The jet has been found before to have a position angle of 116.1\degree{}
\citep{Lee2010HH211,Jhan2021}, here at higher resolution, we can also
pinpoint the closest base of the jet with the high-velocity SiO J=8-7
intensity map to align the jet axis down to the disk scale, as shown in
Figure \ref{fig:cont}c.  As discussed in \cite{Jhan2021}, the high-velocity
ranges of the SiO jet are believed to trace the intrinsic jet coming from
the disk, with the blueshifted velocity range being $\Voff \sim$ $-$34 to
$-$21 \vkm{} and the redshifted velocity range being $\Voff \sim$ 21 to 34
\vkm{}, where the velocity offset $\Voff = \VLSR-\Vsys$ and the systemic
velocity $\Vsys=9.1$ \vkm{}.  The jet has an inclination angle of $\sim$
10\degree{} to the plane of the sky with the blueshifted component in the
southeast (SE) \citep{Jhan2021}.  The closest jet base can then be
determined by the blueshifted component, which is tilted toward us and thus
detected closer to the central protostar than the redshifted component. 
Interestingly the emission roughly forms a conelike structure (although
incomplete) with limb-brightened edges (marked by the dotted lines) opening
to the SE and a tip pointing toward the disk, and thus the tip, which is
marked by an ``x" and roughly coincides with the continuum emission peak,
can be defined as the closest jet base to align the jet axis.  We can apply
the jet axis thus obtained to the redshifted component and find that it can
roughly bisect the redshifted emission at the base.  Notice that the peak of
the redshifted emission located in the southwest (SW) of the jet axis can
trace the edge of the jet, as seen in the blueshifted component.

\subsection{Dusty Disk Structure}

Since the disk is oriented perpendicular to the SiO jet axis, it likely has
a disk axis (rotational axis) aligned with the jet axis and is thus close to
edge-on with the SE side tilted toward us.  As seen in Figure
\ref{fig:cont}d, it is geometrically thick and has a thickness ($\sim$ 20
au) to width ($\sim$ 40 au) aspect ratio of $\sim$ 0.5.  Notice that in
order to better see the disk structure, an unsharp masking filter has been
applied to the continuum map to increase the map contrast.  As indicated by
the three cyan dotted lines, three bright linear structures are seen in the
disk parallel to the equatorial plane, appearing as a 3-layer pancake.  The
SE one and northwest (NW) one are similar to those detected before in
another nearly edge-on but more evolved disk HH 212 \citep{Lee2017Disk}, and
are thus believed to arise from the upper surface and lower surface of the
disk, respectively.  In the HH 212 disk, a thick dark lane is seen in the
midplane.  Such a dark lane is also seen here in the HH 211 disk, but
bisected into two narrow dark lanes by the bright linear structure in the
middle running along the equator, with the SE one broader than the NW one. 
This bright linear structure in the midplane is seen for the first time in a
young disk and could be clumpy.  It extends $\sim$ 16 au to the NE and 24 au
to the SW from the jet axis, and is thus asymmetric about the jet axis. 
Therefore, it is unlikely produced by a circular ring around the central
protostar.  It could be due to an asymmetric structure like a spiral arm
induced by GI to facilitate the disk accretion.  The ``+" sign to the NW of
the brightest peak along the jet axis marks the likely location of the
central protostar, obtained by comparison to the disk model to be discussed
later, and it has a position ICRS $\alpha_{(2000)}=\ra{3}{43}{56}{808}$ and
$\delta_{(2000)}=\dec{32}{00}{50}{1535}$.


\section{Disk Model}





\def\mHt{m_{\textrm{\scriptsize H}_2}}
\def\mH2{m_{\textrm{\scriptsize H}_2}}
\def\no{n_\mathrm{o}}
\def\na{n_\mathrm{t}}
\def\rhoa{\rho_\mathrm{t}}
\def\Ro{R_\mathrm{o}}
\def\zo{z_\mathrm{o}}
\def\Ra{R_\mathrm{t}}
\def\hd{h_\mathrm{d}}
\def\To{T_\mathrm{o}}
\def\Ta{T_\mathrm{t}}
\def\ho{h_\mathrm{o}}
\def\ha{h_\mathrm{t}}
\def\hs{h_\mathrm{s}} 

\def\vk{v_\mathrm{ko}}
\def\cs{c_\mathrm{s}}
\def\vko{v_\mathrm{ko}}
\def\cso{c_\mathrm{so}}
\def\vp{v_\phi}

\def\kabs{\kappa_\textrm{\scriptsize abs}}
\def\ksca{\kappa_\textrm{\scriptsize sca}}


To retrieve the physical properties of the dusty disk, we model the observed
disk emission and structure with a parametrized flared disk model similar to
that used for the HH 212 disk \citep{Lee2021Pdisk}, appropriate for a Keplerian rotating
accretion disk.  
In this model, the dust in the disk is assumed to have the
following mass density and temperature distributions in cylindrical
coordinates $(R,\phi,z)$
\begin{eqnarray} 
\rho (R,z) &=& \rhoa (\frac{R}{\Ra})^{-p}
\exp(-\frac{z^2}{2 \hd^2}) \nonumber \\ 
T (R,z) &=& \Ta (\frac{R}{\Ra})^{-q}
\exp(\frac{z^2}{2 \hd^2})
\end{eqnarray} 
where $\Ra$ is the turnover radius
to be defined below, $\rhoa$ and $\Ta$ are the dust mass density and
temperature in the disk midplane at $\Ra$, respectively.  
The density and
temperature are assumed to decrease with the increasing radius with a
power-law index $p$ and $q$, respectively.
As in the
HH 212 disk \citep{Lee2021Pdisk}, we assume $p=2$ and $q=0.75$.
In addition, the density is assumed to decrease from
the midplane to the surface due to vertical hydrostatic equilibrium
with a scale height $\hd$, while the temperature is assumed to increase from
the midplane to the surface due to the radiative heating by the central
protostar and mechanical heating from the wind-disk interaction.  
The wind-disk interaction is expected because a wide-angle
wind is predicted around the observed jet coming from the innermost disk
\citep{Shu2000,Lee2022Xwind} and a possible disk wind has also been detected
fanning out from the inner disk \citep{Lee2018HH211}.
The scale height depends on the ratio of the sound speed to the angular rotation speed
and thus increases with the
increasing radius.
However, similar to
that of the HH 212 disk, the scale height is assumed to decrease at the turnover
radius $\Ra$ in order to match the observation.
Therefore, we assume
\begin{eqnarray}
\hd (R)= \ha \left\{
\begin{array}{ll} (\frac{R}{\Ra})^{1+(1-q)/2} & \;\;\textrm{if}\;\; R \leq
\Ra, \\ \sqrt{1-\frac{3}{4}(\frac{R-\Ra}{\Ro-\Ra})^2} & \;\;\textrm{if}\;\;
\Ra < R \leq \Ro 
\end{array} \right.  
\label{eq:thick}
\end{eqnarray}
so that the scale height increases to $\ha$ at $\Ra$, and then drops to $\ha/2$ at the outer radius $\Ro$.
In our model, the total height of the disk to calculate the disk emission
is $\ho = \sqrt{2} \hd$.
The disk also contains gas with the same temperature as the dust. The gas
also has the same density distribution as the dust, with
a gas to dust mass ratio of 100.



Radiative transfer assuming LTE is used to calculate the dust emission map
from the model, using the radiative transfer code in \cite{Lee2017Disk}.  A
major uncertainty in the model is the dust absorption opacity.  At the
observed wavelength of 852 \micron{} (i.e, corresponding frequency of 352
GHz), it can be $\kabs \sim 1.9$ cm$^2$ per gram of dust as derived in the
HH 212 disk \citep{Lin2021} in the same evolutionary phase, and be $\kabs
\sim 3.5$ cm$^2$ per gram of dust as extrapolated from that in the T-Tauri
disks \citep{Beckwith1990}.  Here we use the mean value of $\sim$ 2.7 cm$^2$
per gram of dust, with an uncertainty of $\sim$ 30\%.  Dust self-scattering
is ignored because the HH 211 disk is much younger than the HH 212 disk and
dust grain size is expected to be much smaller than 100 \micron{}.  After
the model map is calculated, we sample it on the observed $uv$-coverage, and
then make the simulated map in the same way as the observed map for
comparison.  We also subtract the model visibility from the observed
visibility and then make the residual map to show the goodness of the fit.

\begin{figure}
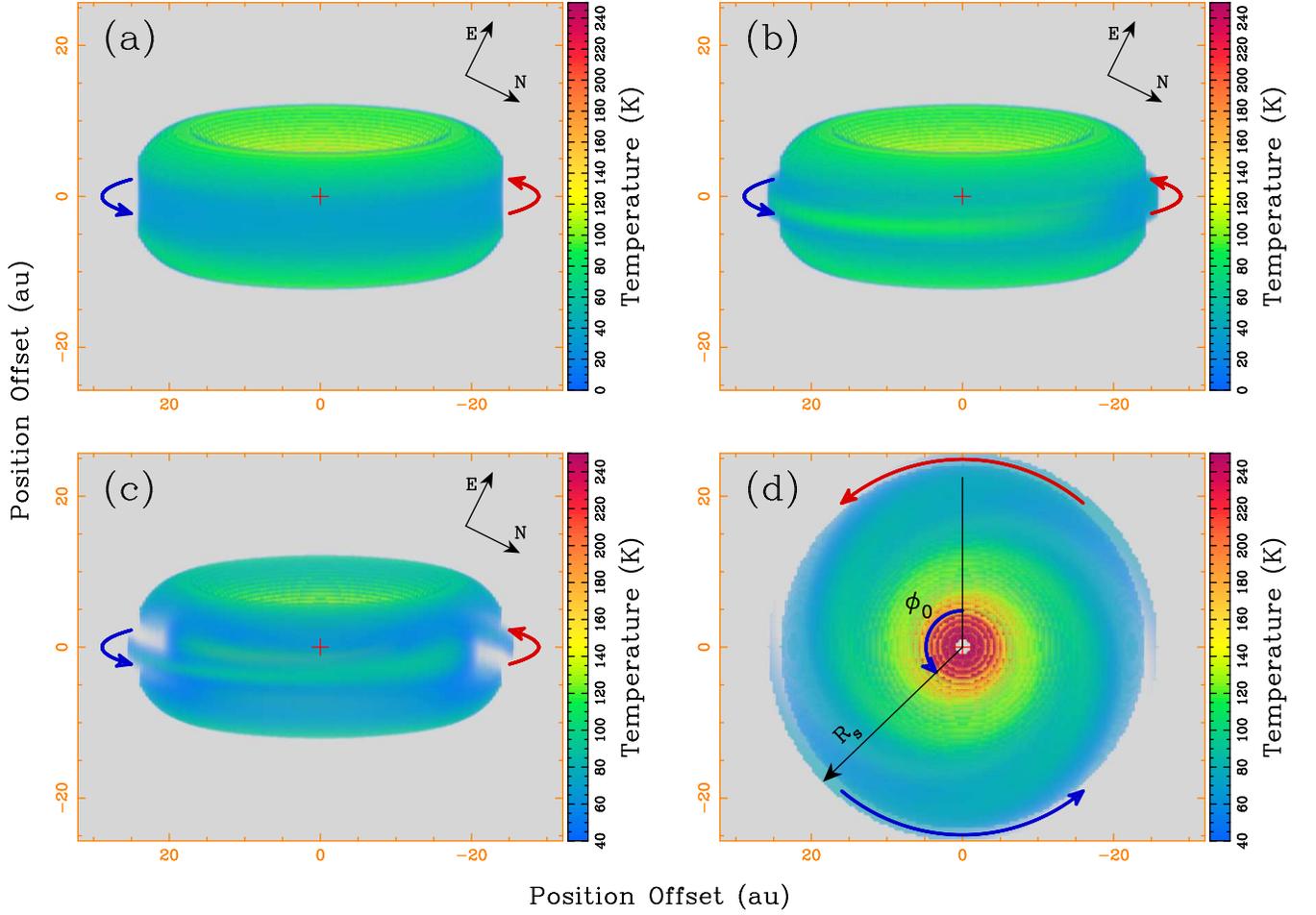

\centering
\putfig{0.7}{270}{f2.eps} 
\figcaption[]
{Disk models to produce the dust emission in the HH 211 disk. 
(a) A disk model without a pair of spiral arms in the midplane of the disk tilted at the observed
inclination angle.
(b)-(d) A disk model with a pair of spiral arms in the midplane. (c) Changed the color range to start from 40 K
in order to reveal the spiral arms in the inner disk. (d) The disk is face-on in order to reveal the spiral arms in the whole disk.
\label{fig:model}}
\end{figure}

Figure \ref{fig:model}a shows the best-fit model (by eye), with $\Ra
\sim$ 18$\pm4$ au, $\Ro \sim 24\pm5$ au, $\ha \sim$ 6$\pm2$ au, $\Ta \sim
40\pm8$ K, and $\rhoa \sim (6.5\pm1.3) \times 10^{-15}$ g \cmc{}.  Here the
uncertainty in geometric parameters is assumed to be 20\% or about a few au,
which is reasonable considering our resolution of $\sim$ 4 au in the disk
minor axis and $\sim$ 7 au in the disk major axis.  The density and
temperature are also assumed to have a 20\% uncertainty, as discussed
below.  The disk is assumed to be slightly tilted, with its (rotational)
axis tilted by $\sim$ 10\degree{} to the plane of the sky as found for the
jet axis \citep{Jhan2021} and with its nearside tilted slightly to the
NW.  With this inclination angle, the upper surface is tilted toward
us, exposing warmer material within the inner radius.  For the lower disk
surface, the inner radius is blocked by the outer disk and thus invisible,
and therefore only the outer radius, which is cooler, is visible.  The
midplane is the coolest.

\begin{figure}
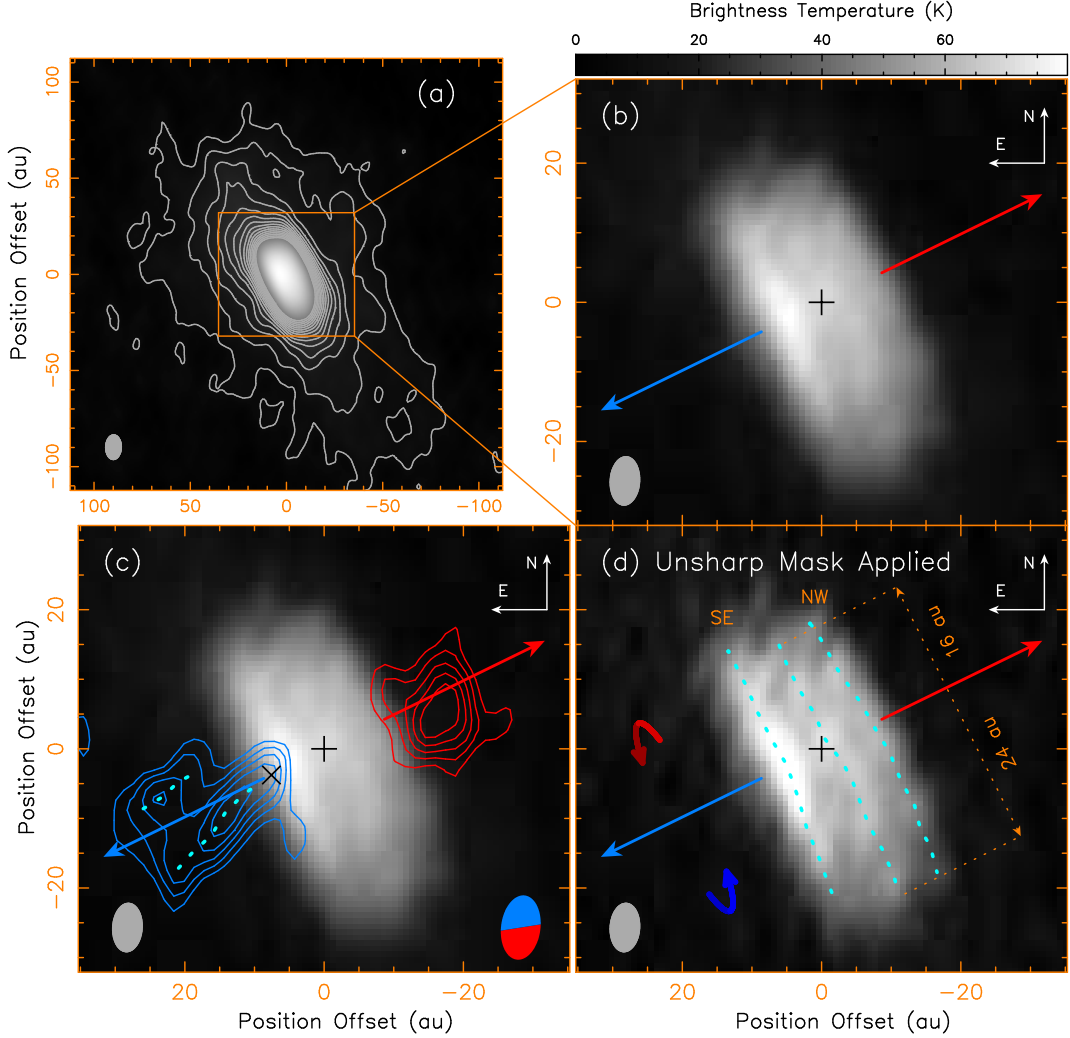

\centering
\putfig{0.75}{270}{f1.eps} 
\figcaption[]
{ALMA continuum maps toward the center of the HH 211 protostellar system at $\sim$ 352 GHz.
The cross marks the position of the central protostar.
The red and blue arrows indicate the axes of the redshifted component
and blueshifted component of the jet axis, respectively. 
Panel (a) shows the continuum map at a resolution of \arcsa{0}{040}$\times$\arcsa{0}{025}.
The contours start at 5$\sigma$ with a step of 4$\sigma$, where $\sigma=0.54$ K.
Panels (b)-(d) show the continuum map at a resolution of \arcsa{0}{022}$\times$\arcsa{0}{013}.
Panel (c) also shows
SiO J=8-7 intensity maps (contours) of the jet at high velocity
in order to determine the jet base (as marked by an ``x") near
the central protostar. The SiO maps have a resolution of \arcsa{0}{029}$\times$\arcsa{0}{018}.
Blue contours show
the blueshifted emission integrated from $\Voff \sim$ $-$34 to $-$21 \vkm{}, while red contours show the redshifted emission
integrated from $\Voff \sim$ 21 to 34 \vkm{}, where $\Voff = \VLSR-\Vsys$, with the systemic velocity $\Vsys=9.1$ \vkm{}. 
The contours start from 3$\sigma$ with a step of 1$\sigma$, where the noise level $\sigma \sim$ 120 K \vkm{}.
The dotted curves show the limb-brightened edges of the jet.
Panel (d) also shows the cyan dotted lines to indicate the three 
linear structures detected in the disk. 
An unsharp masking filter has
been applied to the continuum map to increase the map contrast. The
curved arrows indicate the rotation direction of the disk.
\label{fig:cont}}
\end{figure}

\begin{figure}
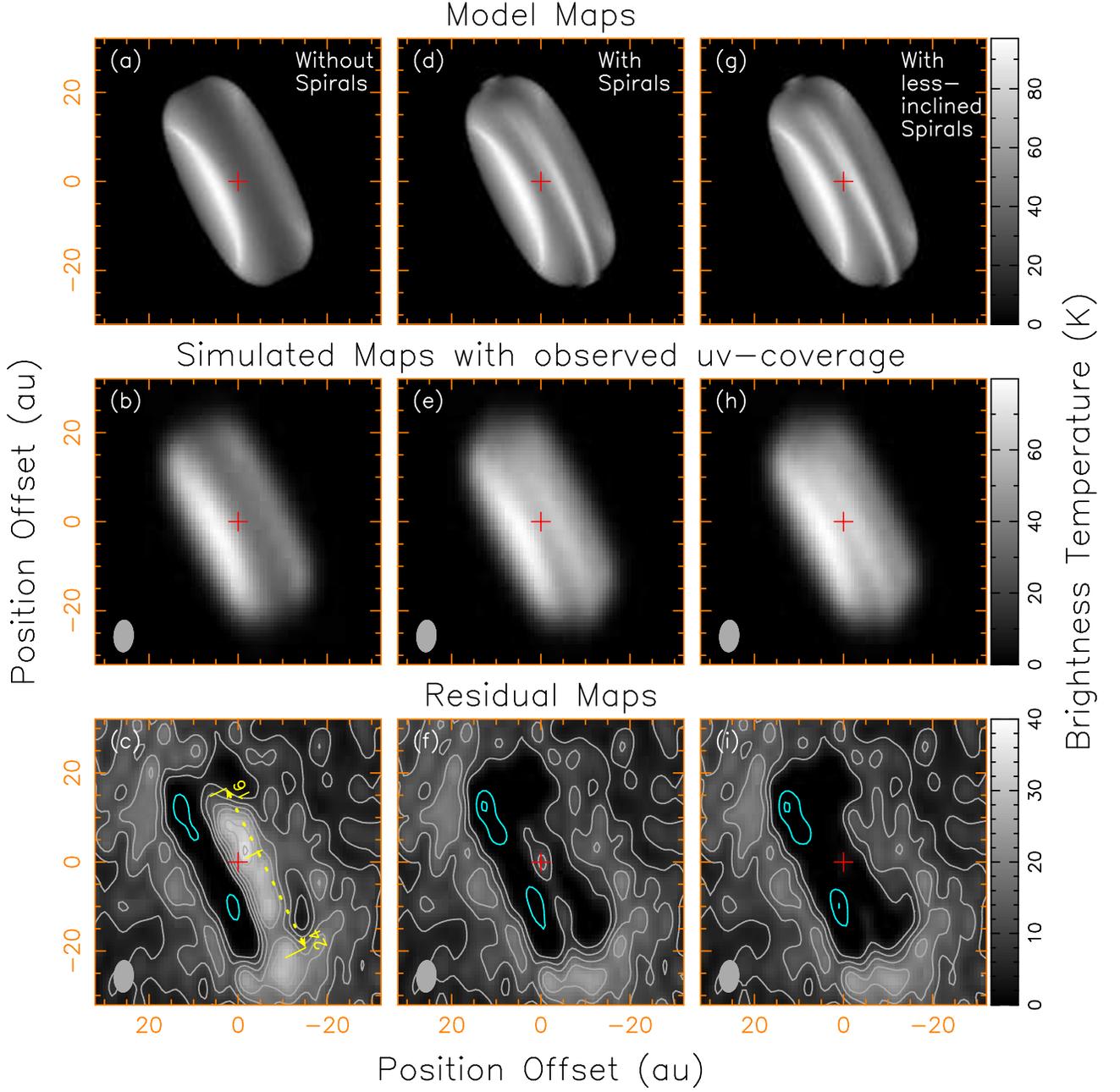

\centering
\putfig{0.9}{270}{f3.eps} 
\figcaption[]
{The model maps, simulated maps, and residual maps from our models. Residual maps are obtained by subtracting
the model visibility from the observed visibility, with
the gray and cyan contours for the positive and negative residuals, respectively.
The contours start from 4$\sigma$ with a step of 3$\sigma$, where $\sigma = 1.6$ K.
Left column shows the maps for the disk model without a pair of trailing spiral arms in the midplane.
Central column shows the maps for the disk model with a pair of trailing spiral arms in the midplane.
Right column shows the maps for the disk model with a pair of slightly less-inclined trailing spiral arms in the midplane.
\label{fig:cont_model}}
\end{figure}

The dust emission map calculated from the disk model with radiative transfer
is shown in Figure \ref{fig:cont_model}a, after rotating by 116.1\degree{}
to match the observed position angle of the disk.  As expected, the map
shows only two bright linear structures arising from the disk surfaces, as
seen in HH 212 \citep{Lee2017Disk}.  Figure \ref{fig:cont_model}b shows the
simulated map sampled with the observed $uv$-coverage and Figure
\ref{fig:cont_model}c shows the residual map obtained by mapping the
residual visibility as mentioned above.  The best-fit model is obtained
by eye by matching the observed structure and emission intensity of
the SE and NW linear structures and thus reducing the residual intensity for
these structures.   Since the geometric parameters $\Ra$, $\Ro$, and
$\ha$ can be readily determined by matching the size and thickness of the disk, we
mainly vary $\Ta$ and $\rhoa$ to match the observed intensity. 
Furthermore, since the disk emission is mostly optically thick, $\Ta$ and
$\rhoa$ can be obtained by matching the observed brightness temperature and
achieving the required optical depth.  As shown in Figure
\ref{fig:cont_model}c, the absolute values of the residuals in the two
linear structures are $\lesssim 13$ K, which is $\sim$ 20\% of the mean
intensity of the observed map of the disk, and thus it is reasonable to
assume a 20\% uncertainty in model density and temperature as mentioned
above. The SE linear structure is from the upper surface and is brighter
because it is from the inner radius exposed to us.  The NW linear structure
is from the lower surface and is fainter because it is from the surface at a
larger radius.  As marked in Figure \ref{fig:cont}, the central protostar is
located closer to the SE linear structure than to the NW linear structure
and is to the NW of the brightest peak along the jet/disk axis, because the
nearside of the disk is tilted slightly to the NW.  The residual map clearly
shows an asymmetric linear structure in the midplane that could not be
fitted by this model.  The linear structure is slightly curved, as expected
for a slightly inclined spiral arm.  After plotting the NE and SW
lengths of the linear structure measured from Figure \ref{fig:cont}d, we
find that the linear structure may extend further to the SW into the
surrounding envelope and observations at higher resolution are needed to
check it. Note that the residual map also shows the emission from the
surrounding envelope that is not included in our model.





Now we add a pair of spiral arms into our model in order to reproduce the
observed linear structure in the midplane because the $m=2$ mode are mostly
seen \citep{Perez2016,Lee2020HH111}.  The spiral arms are assumed to be
trailing, as seen in more evolved disks, e.g., HH 111 \citep{Lee2020HH111}
and Elias 2-26 \citep{Perez2016,Paneque2021}, with their outer tip pointing
in the direction opposite to the disk rotation measured before (as indicated
by the blue and red curved arrows in Figures \ref{fig:cont} and
\ref{fig:model}) \citep{Lee2018HH211}.  
The
spiral arms are assumed to follow the logarithmic structure as excited by GI
due to the self gravity of the disk, with the following amplitude
\begin{equation}
s = \cos[m(\phi-\frac{1}{a}\ln (R/R_s)-\phi_0)].
\end{equation}
Here $m=2$ to produce a pair of spiral arms,
$R_s$ is defined as the outer radius of
the spiral arms at $\phi=\phi_0$ and $\tan^{-1} a$ is the pitch angle.
Then the dust density and temperature
of the disk after adding the spirals are assumed to increase in the vertical direction with the following 
\begin{eqnarray}
\rho^s(R,\phi,z) &=& \rhoa (\frac{R}{\Ra})^{-p} [\exp(-\frac{z^2}{2 \hd^2})+A(1+s) \exp(-\frac{z^2}{2 \hs^2})] \nonumber \\
   T^s(R,\phi,z) &=& \Ta (\frac{R}{\Ra})^{-q}[\exp(\frac{z^2}{2 \hd^2})+ B(1+s) \exp(-\frac{z^2}{2 \hs^2})] \nonumber \\
                 & &
\end{eqnarray}
where the scale height of the spiral arms is assumed to be a fraction of the disk scale height,
i.e., $\hs = f \hd$. $A$ and $B$ are the spiral amplitude in density and temperature, respectively.
Thus, in the midplane, the disk density and temperature have a variation amplitude of $A/(1+A)$ and
$B/(1+B)$, respectively.

Figure \ref{fig:model}b shows the best-fit model (by eye) at the
observed inclination angle, revealing the spiral arm in the front.  The
spiral arm in the back is blocked by the disk.  Figure \ref{fig:model}c
shows the spiral arms further in by changing the colorbar range to hide the
material with a temperature lower than 40 K.  Figure \ref{fig:model}d shows
the face-on view in order to view the spiral arms from the top.  As seen in
the resulting model map in Figure \ref{fig:cont_model}d and simulated map in
Figure \ref{fig:cont_model}e, the spiral arm in the outer disk produces a
linear structure in the midplane.  It is trailing, and thus reproducing the
observed asymmetry about the jet axis.  In the model, the geometric
parameters of the spiral arms $R_s$, $\phi_0$, and $a$ can be determined by
the NE and SW lengths of the linear structure in the midplane, assuming that
the linear structure comes from the spiral arm in the front part of the
outer disk.  With $\phi_0 \sim 135$\degree{} (i.e., $\frac{3}{4}\pi$) and
$R_s \sim 26$ au, the model produces a linear structure with a SW length of
$\sim$ 24 au, as observed.  Then in order to account for the length of
$R_i\sim 16$ au in the NE where $\phi_i=270$\degree{} (i.e., $\frac{3}{2}\pi$),
we have $a = \ln(R_i/R_s)/(\phi_i-\phi_0) \sim -0.2$ (negative for
trailing). Thus, the pitch angle is $\tan^{-1} |a| \sim 11$\degree{}, which
is not much different from those of other spiral arms found in more evolved
disks, e.g., HH 111 (13\degree{}$-$16\degree{}) \citep{Lee2020HH111} and
Elias 2-27 ($\sim$ 16\degree{}) \citep{Perez2016}.   Notice that the
outer radius and thus the pitch angle of the spiral arms can be larger if
the observed linear structure turns out to extend further to the SW into the
envelope, as discussed earlier.

By roughly matching the observed intensity and thickness of the linear
structure in the midplane, we find that $A\sim 1$, $B\sim 1$, and $f \sim
0.3$. Our fitting results indicate that the disk density and temperature
need a variation of $\sim$ 50\% in the midplane from the spiral arm region
to non-spiral arm region.  The material is required to be denser and warmer
within the spiral arms than the original disk to match the observed
intensity.  The spiral arm comes from a radius of $\sim$ 15 to 26 au, where
it has a midplane temperature of 90$-$130 K.  Density enhancement is
required there to provide the necessary optical depth of $\sim$ 1 for the
spiral arm to be detected at the observed brightness temperature of $\sim$
50$-$70 K.  Also, the scale height of the spiral arms is found to be $\sim$
30\% of that of the disk in order to roughly match the thickness of the
observed linear structure.  Since the observed spiral arm in the midplane is
not spatially well resolved in our observations, the scale height of the
spiral arm could be smaller, and thus the temperature could be higher.




Nonetheless, the residual map (Figure \ref{fig:cont_model}f) still
shows a faint and narrow residual emission structure on the SE edge of the
spiral arm, indicating that the spiral arm should be slightly less inclined
than the disk.  Assuming that the spiral arm is slightly less inclined by
$\sim$ 2\degree{} than the disk, this residual emission structure disappears
in the residual map, as seen in Figure \ref{fig:cont_model}i.  In this case,
the linear structure produced by the spiral arm (see Figures
\ref{fig:cont_model}g and \ref{fig:cont_model}h) can also better bisect the
dark lane into two narrower dark lanes, with the SE one slightly broader
than the NW one, as observed.





\begin{figure}
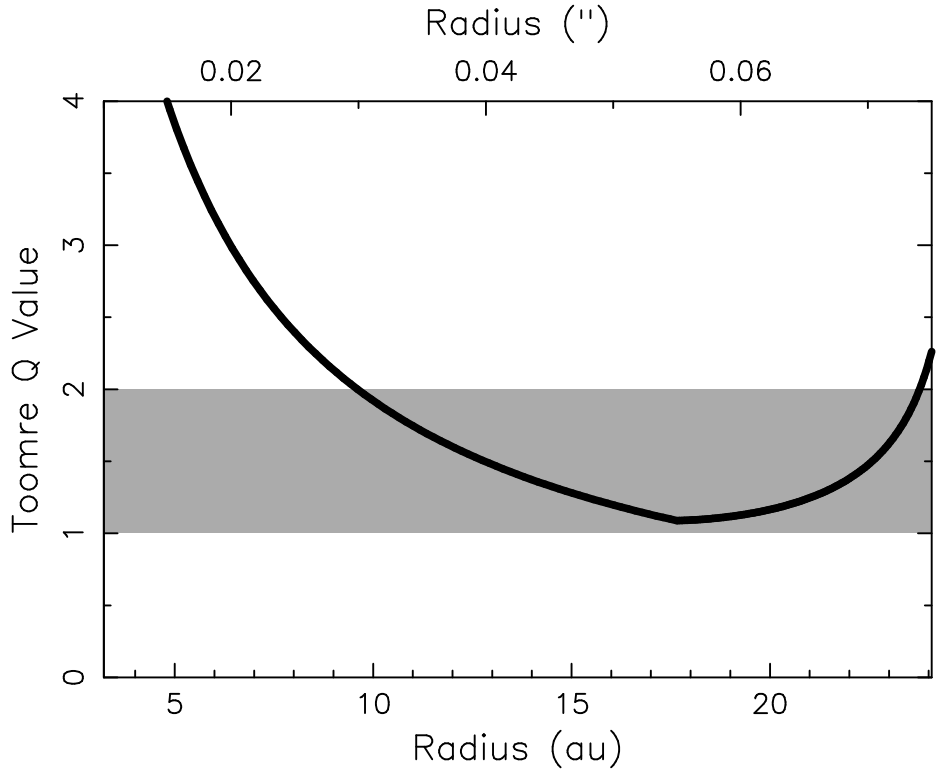

\centering
\putfig{1}{270}{f4.eps} 
\figcaption
{Toomre Q values for our simple model with a pair of trailing spiral arms in the midplane.
\label{fig:Qvalue}}
\end{figure}

In summary, the observed structure and emission intensity of the HH 211 disk
can be roughly reproduced by our simple disk model with warm disk surfaces
and a warm trailing spiral arm in the midplane.  In our model, the disk has
a mass of $M_d\sim$ 0.045 \solarmass{} (see Appendix).  Since the total
mass of the disk and protostar has been estimated before to be $M_t\sim$
0.08 \solarmass{} from the kinematic study of the disk and envelope
\citep{Lee2019HH211}, the protostar itself has a mass of $M_\ast = M_t-M_d
\sim $ 0.035 \solarmass{}.
 Hence, the protostar currently has
a mass of a brown dwarf and is expected to grow later into a low-mass star
with a mass of $\sim$ 0.3$-$0.8 \solarmass{} \citep{Froebrich2005}. 
Comparable disk and protostellar masses are also seen in numerical
simulations at early evolutionary stages ($<0.1$ Myr) when the protostellar
mass is low ($<0.1 - 0.2$ \solarmass{}) \citep{Tsukamoto2020}.  In this
case, GI develops, producing spiral arms transporting angular momentum
within the disk.  The observed linear structure in the midplane could trace
such a spiral arm, especially when the resulting Toomre Q values in our
model's outer disk are $<2$ (see Figure \ref{fig:Qvalue} and Appendix for
the derivation) and thus optimal for GI.  The high envelope infall rate of
$\sim$ 4 $\times10^{-6}$ \solarmasspy{} \citep{Lee2019HH211} onto the disk
also supports this possibility \citep{Tomida2017}.  In addition, the disk to
protostar mass ratio is $\eta = M_d/M_\ast\sim 1.3$ and thus GI is expected to induce
prominent $m=2$ spiral arms \citep{Dong2015,Tsukamoto2020}, as assumed in
our model.   Nonetheless, higher resolution observations are needed
to check if $m \neq 2$ can also reproduce the linear structure in the midplane.
The spiral arms have a scale height about 30\% of the disk scale
height, indicating that they start in the midplane where the density is
high.


\section{Discussion}

Spiral arms have been detected in actively accreting but more evolved
systems when the disks are large ($\gtrsim 100$ au in radius), e.g., HH 111
\citep{Lee2020HH111}, Elias 2-27 \citep{Perez2016,Paneque2021}, and possibly
TMC-1A \citep{Aso2021}.  These spiral arms are trailing and can be induced
by GI to transport angular momentum to facilitate the disk accretion
\citep{Kratter2016,Tomida2017}.  Unlike the HH 211 disk, these larger more
evolved disks have settled more or less to a thin disk geometry.  In the
earlier phase of star formation when the disks are still geometrically
thick, hints of spiral arms have also been detected in, e.g., HH 212
\citep{Lee2021Pdisk} and L1527 \citep{Ohashi2022,Sheehan2022}.  Now in the
earliest phase of the HH 211 disk, which is geometrically thicker, our
observations and modeling have strongly supported a presence of a trailing
spiral arm in the midplane.  Our result in the HH 211 disk urges us to
confirm those previous hints of spiral arms, in order to determine if
actively accreting disks can always be gravitational unstable in order to
induce spiral arms to transport angular momentum, as seen in many
simulations \citep{Forgan2018,Tomida2017,Tsukamoto2020}.  In simulations,
when cooling is efficient, spiral arms can fragment into clumps that can
grow into planets \citep{Gammie2001,Rice2003,Forgan2018Frag}.  Hence, further
work is needed to investigate the origin of the clumps seen in the HH 211
disk and if the clumps can grow into planets.


The HH 211 disk is geometrically thick, a trait also seen in the more
evolved and also vertically resolved edge-on disks, e.g., HH 212
\citep{Lee2017Disk} and L1527 \citep{Nakatani2020}.  It is flared and likely
in vertical hydrostatic equilibrium, with the scale height roughly
consistent with that derived from the ratio of the sound speed to the
angular rotation speed, as discussed in Appendix.  This indicates that the
submillimeter light emitting dust grains have yet to settle to the midplane. 
This is different from the protoplanetary disks, which are very thin in dust
continuum images of ALMA \citep{Brogan2015,Villenave2020} likely because the
dust grains have grown larger and settled to the midplane
\citep{Villenave2020}.  Moreover, the disk in the early phase is more
turbulent because it has a higher accretion rate, spiral arms perturbation
induced by GI, and more energetic wind-disk interaction due to a higher
mass-loss rate, possibly preventing the dust from settling to the midplane.




Our ALMA observations of the HH 211 disk also provide a strong constraint on
the magnetic braking effect on the disk formation.  Previous studies in this
system have revealed a collapsing (infalling) and rotating flattened
envelope with a centrifugal barrier at a radius of $\sim$ 20 au
\citep{Lee2019HH211}.  This radius coincides with the outer radius of the
disk resolved here, supporting the formation of the disk around the
centrifugal barrier, without significant loss of angular momentum.  This
could be because the rotational axis is largely misaligned with the magnetic
field axis in the collapsing core on the larger scale \citep{Lee2019HH211},
which can reduce the magnetic braking in the envelope \citep{Joos2012}. 
Another possible explanation could be because of the ambipolar diffusion
that increases the mass-to-flux ratio and thus reduces the magnetic braking
in the envelope \citep{Yen2023}.  In addition, since the material
accumulates around the centrifugal barrier, the Toomre Q values are expected
to decrease there, triggering GI and thus forming spiral arms, transferring
material inward.

\acknowledgements

We thank the anonymous referee for useful comments.
This paper makes use of the following ALMA data:
ADS/JAO.ALMA\#2019.1.00570.S.  ALMA is a partnership of
ESO (representing its member states), NSF (USA) and NINS (Japan), together
with NRC (Canada), NSC and ASIAA (Taiwan), and KASI (Republic of Korea), in
cooperation with the Republic of Chile.  The Joint ALMA Observatory is
operated by ESO, AUI/NRAO and NAOJ.  C.-F.L. and K.S. Jhan
acknowledge grants from the Ministry of Science and Technology of Taiwan
(MoST 107-2119-M-001-040-MY3, 110-2112-M-001- 021-MY3) and the Academia
Sinica (Investigator Award AS-IA-108-M01).
















\clearpage

\appendix

\section{Extended Properties of the Disk Model}


\def\xo{x_\textrm{\scriptsize 0}}

Using the parametrized flared dusty disk with spiral arms model given in the main body, we can derive
the extended properties of the disk.
The disk has a surface density of dust and gas given by
\begin{eqnarray}
\Sigma (R) &\sim& 100 \int_{-\ho}^{\ho} \int_0^{2\pi} \rho^s (R,\phi,z) \frac{d\phi}{2\pi} \,dz \nonumber \\
&= & 100 \int_{-\ho}^{\ho} \rhoa (\frac{R}{\Ra})^{-p}\Big[\exp(-\frac{z^2}{2 \hd^2})+A \exp(-\frac{z^2}{2 \hs^2})\Big] \,dz \nonumber \\
&= & 100 \sqrt{2 \pi} \rhoa\, (\frac{R}{\Ra})^{-p} \; \hd \Big[\textrm{Erf}(1)+ A f \Big]\nonumber \\
& =& \Sigma_t \left\{
\begin{array}{ll} (\frac{R}{\Ra})^{1+(1-q)/2-p} & \;\;\textrm{if}\;\; R \leq
\Ra, \\ (\frac{R}{\Ra})^{-p} \sqrt{1-\frac{3}{4}(\frac{R-\Ra}{\Ro-\Ra})^2} & \;\;\textrm{if}\;\;
\Ra < R \leq \Ro 
\end{array} \right. 
\end{eqnarray}
with the surface density of dust and gas at $\Ra$ given by
\begin{equation}
\Sigma_t \equiv 100 \sqrt{2 \pi} \Big[\textrm{Erf}(1)+ A f \Big] \rhoa \ha \,.
\end{equation} 
Then, the disk has a mass of
\begin{equation}
M_d  =  \int_0^{\Ro} \Sigma(R)\, 2 \,\pi \,R dR =  2 \pi C \,\Ra^2 \Sigma_t
\end{equation}
with
\begin{equation}
C \equiv \frac{2}{7-2p-q}+\int_1^{\xo}
x^{-p+1} \sqrt{1-\frac{3}{4}(\frac{x-1}{\xo-1})^2} dx
\end{equation}
where $\xo \equiv \frac{\Ro}{\Ra}$. The Toomre Q value can then be given by
\begin{equation}
Q (R) \approx \frac{c_s \Omega}{\pi G \Sigma(R)}
\end{equation}
where $\Omega$ is the angular rotation speed of the disk and $c_s$ is the sound speed
\begin{equation}
c_s =\sqrt{\frac{\gamma kT}{\mu m_H}}
\label{eq:cs}
\end{equation}
where $\mu=2.33$ and $\gamma=7/5$ for molecular gas with H$_2$ and Helium.
Since we are more interested in the outer disk around $\Ra$, we assume
\begin{equation}
\Omega =\sqrt{ \frac{G M_t}{R^3}}
\end{equation}
with $M_t$ being the total mass within $\Ra$ and given by
\begin{equation}
M_t = M_\ast+  \frac{4\pi}{7-2p-q} \Ra^2 \Sigma_t 
\end{equation}
where $M_\ast$ is the protostellar mass.
Since the temperature increases from the midplane to the surface, we use
the density weighted temperature averaged over the disk height, i.e.,
\begin{eqnarray} 
\bar{T}(R) &=&  \frac{\int \int \rho^s(R,z) T^s(R,z) \frac{d\phi}{2\pi}dz}{\int \int \rho^s(R,\phi, z) \frac{d\phi}{2\pi} dz} \nonumber \\
&\approx& \Ta (\frac{R}{\Ra})^{-q} \frac{2 \sqrt{2} + (A+B) \sqrt{2 \pi} f + A\,B \sqrt{\pi} f}
{\sqrt{2 \pi} \,\textrm{Erf}(1)+A \sqrt{2 \pi} f }
\end{eqnarray}
to calculate the value of $c_s$ at $R$ with Eq. \ref{eq:cs}
and then the value of $Q(R)$.

With our best-fit parameters, $\Sigma_t \sim$ 180 g \cms{} and then the disk
has a (gas and dust) mass of $M_d \sim 0.045$ \solarmass{}.  This in
turn implies that the central protostar has a mass of $M_\ast \sim$ 0.035
\solarmass{}, because the mass of disk and protostar together has been
estimated to be $\sim$ 0.08 \solarmass{} from the kinematic study of the
disk and envelope \citep{Lee2019HH211}.    Figure \ref{fig:Qvalue} presents the
resulting Toomre-Q values, which are $< 2$ in the outer disk, optimal
for GI.  With the density weighted temperature, the theoretical scale height
is $\sim c_s/\Omega \sim$ 5.5 au at $\Ra$, consistent with the best-fit
scale height $\ha$, supporting that the disk is close to be in vertical
hydrostatic equilibrium.








\end{document}